\documentclass{llncs}
\usepackage{amsmath, amssymb, graphics}
\usepackage{fullpage}
\usepackage{graphicx}

\newcommand{\Prob}[1]{\mathbf{P} \left( #1 \right)}
\newcommand{\Expec}[1]{\mathbf{E} \left[ #1 \right]}
\newcommand{\ideaproof}{\noindent\textit{Idea of the proof. }}

\begin{document}

\title{A Note on Uniform Power Connectivity in the SINR Model\thanks{Zvi Lotker and Francesco Pasquale were partially supported by a gift from Cisco Reseach Center}}
\author{Chen Avin$^1$, Zvi Lotker$^1$, Francesco Pasquale$^2$, Yvonne-Anne Pignolet$^3$}
\institute{$^1$Ben Gurion University of the Negev, Israel \; $^2$University of Salerno, Italy \; $^3$ETH Zurich, Switzerland\\
$\{$avin,zvilo$\}$@cse.bgu.ac.il, pasquale@dia.unisa.it, pignolet@tik.ee.ethz.ch}
\maketitle

\begin{abstract}
In this paper we study the connectivity problem for wireless networks under the Signal to Interference plus Noise Ratio (SINR) model.
Given a set of radio transmitters distributed in some area, we seek to build a directed strongly connected communication graph, and compute an edge coloring of this graph such that the transmitter-receiver pairs in each  color class can communicate simultaneously. Depending on the interference model, more or less colors, corresponding to the number of frequencies or time slots, are necessary. We consider the SINR model that compares the received power of a signal at a receiver to the sum of the strength of other signals plus ambient noise . The strength of a signal is assumed to fade polynomially with the distance from the sender, depending on the so-called path-loss exponent $\alpha$.

We show that, when all transmitters  use the same power, the number of colors needed is constant in one-dimensional grids if $\alpha>1$ as well as in two-dimensional grids if $\alpha>2$. For smaller path-loss exponents and two-dimensional grids we prove upper and lower bounds in the order of $\mathcal{O}(\log n)$ and $\Omega(\log n/\log\log n)$ for $\alpha=2$ and $\Theta(n^{2/\alpha-1})$ for $\alpha<2$ respectively. If nodes are distributed uniformly at random on the interval $[0,1]$, a \emph{regular} coloring of $\mathcal{O}(\log n)$ colors guarantees connectivity, while $\Omega(\log \log n)$ colors are required for any coloring.
\end{abstract}

\section{Introduction}
The performance of wireless networks depends on the coordination of the timing and frequency bands of broadcasting nodes. This is due to the fact that if two nodes close to each other transmit concurrently, the chances are that neither of their signals can be received correctly because of interference. Thus, choosing an appropriate interference model is critical. The most popular models can be divided into two classes:  graph-based models (protocol models) and  fading channel models. Graph-based models, such as the unit disk graph (UDG) model \cite{Clark1990unit}, describe interference as a binary property by a set of interference edges. The existence of an edge between two communication pairs, usually based on the distance between nodes, implies that the two pairs cannot transmit successfully at the same time (or on the same frequency). Such models, serving as a simple abstraction of wireless networks, have been very useful for the design of efficient distributed algorithms. Nevertheless, graph-based models bear the limitation of representing interference as a local property. In reality, the interference of several concurrent senders accumulates and can interrupt the reception at a far-away receiver. Therefore, the focus of the algorithmic networking community has recently shifted from graph-based models to the more realistic fading channel models, such as the physical Signal to Noise plus Interference (SINR) model \cite{kumar00} that we use in this paper. In this model, a message is received successfully if the ratio between the strength of the sender signal at the receiving location and the sum of interferences created by all other simultaneous senders plus ambient noise is larger than some hardware-defined threshold. Interference is modeled as a continuous property, decreasing polynomially with the distance from the sender, according to the value of the so-called path-loss exponent $\alpha$. More formally, a receiver $r_i$ receives a sender $s_i$'s transmission if and only if
$$
\frac{\frac{P(s_i)}{d(s_i,r_i)^\alpha}}{N+\sum_{j\neq i}{\frac{P(s_j)}{d(s_j,r_i)^\alpha}}} \;\geqslant \; \beta,
$$
where $P(s_k)$ denotes the transmission power of sender $s_k$, $d(s_k,r_i)$ is the distance between sender $s_k$ and receiver $r_i$, $N$ denotes the ambient noise power level and $\beta$ is the minimum SINR required for the successful reception of a message.

\begin{figure}[t]
\centering
\begin{tabular}{ccc}
\begin{minipage}{1.8in}
\centering
\includegraphics[width=1.8in]{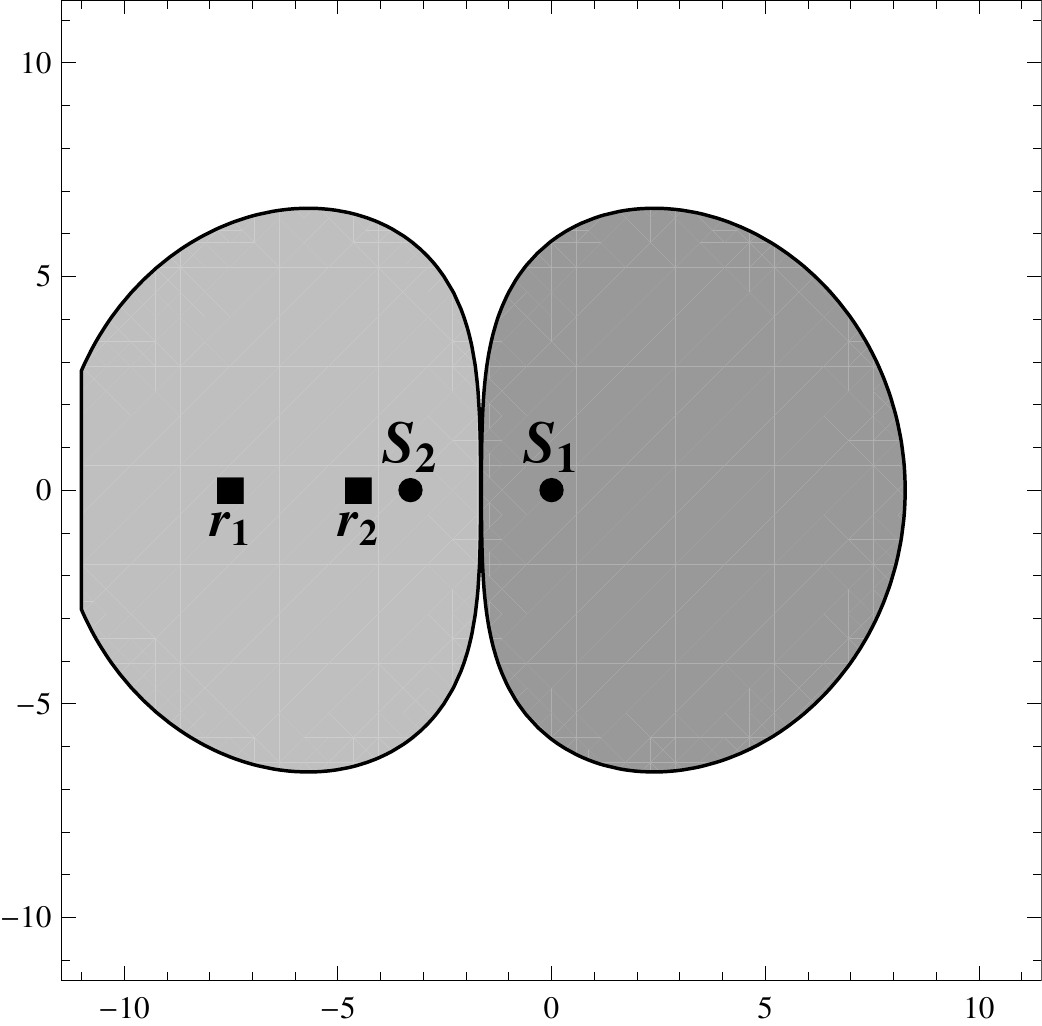}
\end{minipage}& &
\begin{minipage}{1.8in}
\centering
\includegraphics[width=1.8in]{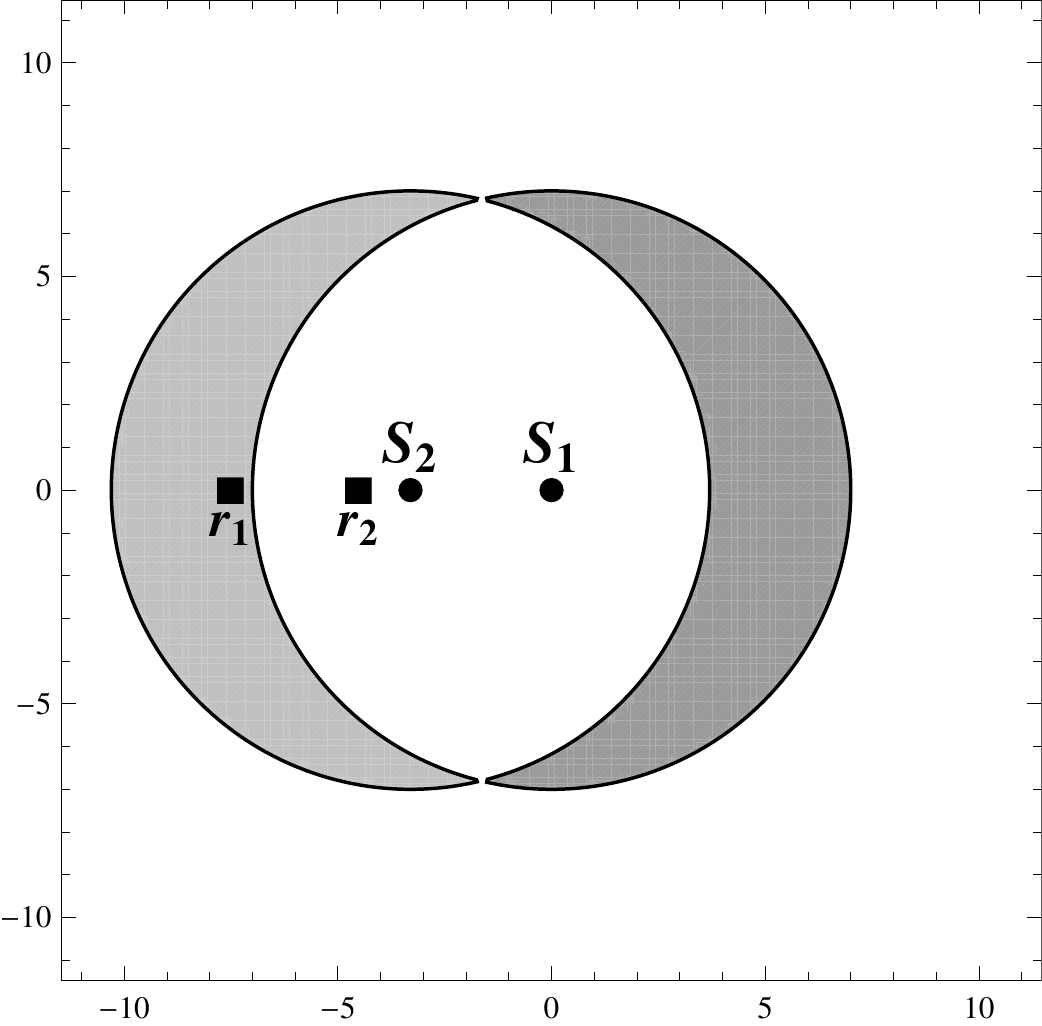}
\end{minipage}\\
\small (a) & & \small (b)\\
\end{tabular}
\caption{Reception diagrams for scenario with two links, $l_1 = (s_1,r_1)$ and $l_2 = (s_2, r_2)$. The shaded areas denote where the signal of a sender can be decoded (the area in the lighter gray belongs to sender $s_2$), white indicates that the received signal power is too weak for reception.
\emph{(a)} SINR model: only node $r_2$ receives a message from its sender, the interference is too high at $r_1$. \emph{(b)} Unit Disk Graph model:
neither $r_1$ nor $r_2$ receive a message from their corresponding senders.}
\label{fig:uniform}
\end{figure}

In this paper we focus on the uniform power assignment, where every node transmits with the same power. This strategy has several important advantages due to its simplicity. While the benefits of power control are obvious, wireless devices that always transmit at the same power are less expensive and less complicated to build. Therefore, the uniform power assignment has been widely adopted in practical systems.
From the algorithmic perspective, the lack of freedom in choosing power levels makes reaching a decision much simpler. Moreover, recently a study of
\emph{SINR diagrams}\footnote{The SINR diagram of a set of transmitters divides the plane into $n+1$ \emph{regions} or reception zones, one region for each transmitter that indicates the set of locations in which it can be heard successfully, and one more region that indicates the set of locations in which no sender can be heard. This concept is perhaps analogous to the role played by Voronoi diagrams in computational geometry.}
~\cite{Avin2008SINR}  showed that the reception zones of all senders are \emph{convex} for a uniform scheme but not necessarily for non-uniform power
assignments. This finding suggests  that designing algorithms may be much simpler for uniform networks than for non-uniform networks.
Figure 1 illustrates a setting with uniform power levels in the SINR and in the UDG model.

In any network, it is typically required that any pair of nodes can exchange message via relay nodes. In other words, the nodes have to be connected by a communication backbone, e.g., a spanning tree or a connected dominating set. In this paper, we investigate how many colors (time slots / frequencies) are necessary to guarantee that the resulting links (node pairs that can communicate) form a connected graph. \cite{Moscibroda2006Complexity} was the first to explore this question in the physical interference model. The authors suggest an algorithm that constructs a spanning tree, and assigns power levels and time slots to each link of the tree. This algorithm guarantees that at most $\mathcal{O}(\log^4 n)$ colors suffice for all transmissions to be received correctly, i.e., even in worse-case networks, the scheduling complexity of a strongly-connected topology is polylogarithmic in $n$ and such topologies can thus be scheduled efficiently. The algorithm assigns many different power levels to the links and does not lend itself to a distributed implementation. As we discussed earlier, the study of the uniform case is still worthwhile, thanks to its simplicity and the way cheap commercial hardware is built. Therefore we aim at shedding light on the connectivity problem for uniform power assignments in this paper. More precisely, given a coloring we can construct a \emph{SINR graph}, that represents which nodes can communicate concurrently. We examine the number of colors are necessary such that a strongly connected SINR graph can be built. We show that the number of colors needed is constant in one-dimensional grids if $\alpha>1$ as well as in two-dimensional grids if $\alpha>2$. For smaller path-loss exponents, more colors are necessary. If $\alpha=2$ (i.e., the signal propagation in the vacuum), the upper and lower bounds for the number of colors are in the order of $\mathcal{O}(\log n)$ and $\Omega(\log n/\log\log n)$ respectively. Even smaller values of $\alpha$ have been measured for indoor propagation \cite{rappaport}. For $\alpha<2$ we provide a tight bound of $\Theta(n^{2/\alpha-1})$. For the special case of $\alpha=2$ we examined the connectivity of nodes distributed uniformly at random on the interval [0,1]. In this setting, a regular coloring of $\Theta(\log n)$ colors guarantees connectivity.

\section{Related Work}
The seminal work of Gupta and Kumar~\cite{kumar00} initiated the study of the capacity of wireless networks. The authors bounded the throughput capacity in the best-case (i.e., optimal configurations) for the protocol and the physical models for $\alpha > 2$.

For both model classes, many scheduling algorithms have been suggested. E.g., \cite{HS88,KMPS04,ShMaSh06} analyze algorithms in graph-based models. Typically, these algorithms employ a coloring strategy, which neglects the aggregated interference of nodes located further away. The resulting inefficiency of graph-based scheduling protocols in practice is well documented, both theoretically and by simulation \cite{gro05,MoWa06} as well as experimentally \cite{Moscibroda2006Protocol}. Recently, Lebhar et al. \cite{Lebhar2009Unit} consider the case of $\alpha>2$  and senders that are deployed uniformly at random in the area. They showed how a UDG protocol can be emulated when the network operates under the SINR model. Their emulation cost factor is $\mathcal{O}(\log^3 n)$. The fact that interference is continuous and accumulative as well as the geometric constraints lead to an increased difficulty of the scheduling task in the SINR model, even if the transmission power of the nodes is fixed. Two scheduling problems are shown to be NP-complete in the physical SINR model in \cite{GOW07}. Goussevskaia et al. propose in \cite{GHWW09} a scheduling algorithm with an approximation guarantee independent of the network's topology. Their algorithm gives a constant approximation for the problem of maximizing the number of simultaneously feasible links and leads to a $\mathcal{O}(\log n)$ approximation for the problem of minimizing the number of time slots to schedule a given set of requests. Furthermore, in~\cite{Halld'orsson2009}, the problem is shown to be in APX, thus precluding a PTAS and the authors propose an improved algorithm leading to a constant approximation. Yet another line of research investigates static properties under the SINR model, e.g., the maximum achievable signal-to-interference-plus-noise ratio~\cite{Zander92b} or the shape of reception zones of nodes in a network \cite{Avin2008SINR}.

Non-uniform power assignments can clearly outperform a uniform assignment \cite{Moscibroda2006Protocol,MoWa06} and increase the capacity of the network, therefore the majority of the work on capacity and scheduling addressed non-uniform power. Recent work \cite{Avin09} compares the uniform power assignment with power control when the area where nodes, whereas \cite{Fanghanel2009,Halld'orsson09b} give upper and lower bounds for power-controlled oblivious scheduling. As mentioned in the introduction, Moscibroda et al. \cite{Moscibroda2006Complexity} were the first to raise the question of the complexity of connectivity in the SINR model. While their work applies for networks with devices that can adjust their transmission power, we address networks composed of devices that transmit with the same power.

\section{Model}
Let $(M,d)$ be a metric space and $V \subseteq M$ a finite set of $n=|V|$ \emph{nodes}. A node $v_j$ successfully receives a message from node $v_i$ depending on the set of concurrently transmitting nodes and the applied interference model. A standard interference model that captures some of the key characteristics of wireless communication and is sufficiently concise for rigorous reasoning is the physical SINR model~\cite{kumar00}. In this model, the successful reception of a transmission depends on the strength of the received signal, the interference caused by nodes transmitting simultaneously, and the ambient noise level. The received power $P_{r_i}(s_i)$ of a signal transmitted by a sender $s_i$ at an intended receiver $r_i$ is $P_{r_i}(s_i)\;=\;{P(s_i)}\cdot{g(s_i,r_i)},$ where $P(s_i)$ is the transmission power of $s_i$ and $g(s_i,r_i)$ is the propagation attenuation (link gain) modeled as $g(s_i,r_i)=d(s_i,r_i)^{-\alpha}$. The \emph{path-loss exponent} $\alpha \geqslant 1$ is a constant typically between $1.6$ and $6$. The exact value of $\alpha$ depends on external conditions of the medium (humidity, obstacles, etc.) and on the exact sender-receiver distance. Measurements for indoor and outdoor path-loss exponents can be found in \cite{rappaport}.

Given a sender and a receiver pair $l_i = (s_i, r_i)$, we use the notation $I_{r_i}(s_j)=P_{r_i}(s_j)$ for any other sender $s_j$ concurrent to
$s_i$ in order to emphasize that the signal power transmitted by $s_j$ is perceived at $r_i$ as interference. The \emph{total interference} $I_{r_i}(L)$ experienced by a receiver $r_i$ is the sum of the interference power values created by the set $L$ of nodes transmitting simultaneously on the same frequency (except the intending sender $s_i$), i.e., , $I_{r_i}(L):=\sum_{l_j\in L\setminus \{l_i \}}{I_{r_i}(s_j)}$. Finally, let $N$ denote the ambient noise power level. Then, $r_i$ receives $s_i$'s transmission if and only if
$$
\text{SINR}(l_i)=\frac{P_{r_i}(s_i)}{N+I_{r_i}(L)}
=\frac{{P(s_i)}g(s_i,r_i)}{N+\sum_{
j\neq i}{{P(s_j)}{g(s_j,r_i)}}}
=\frac{\frac{P(s_i)}{d(s_i,r_i)^\alpha}}{N+\sum_{
j\neq i}{\frac{P(s_j)}{d(s_j,r_i)^\alpha}}} \;\geqslant \; \beta,
$$
where $\beta \geqslant 1$ is the minimum SINR required for a successful message reception. For the sake of simplicity, we set $N=0$ and ignore the influence of noise in the calculation of the SINR. However, this has no significant effect on the results: by scaling the power of all senders, the influence of ambience noise can be made arbitrarily small. Observe that for real scenarios with upper bounds on the maximum transmission power this is not possible,  however, for our asymptotic calculations we can neglect this term. We assume that every node can listen/send on all available frequencies simultaneously.

The \emph{scheduling complexity}, introduced in \cite{Moscibroda2006Complexity}, describes the number of time slots or frequencies necessary to successfully transmit messages over a given set of communication links. More formally, we are given a network with a set of directed links representing communication requests. For each such link we assign a color (time slot/frequency) and a power level such that all simultaneous transmissions are successful, i.e., not violating the signal-to-interference plus noise ratio at any receiver.

The \emph{connectivity problem} of a given set $V$ of nodes located in the Euclidean plane is the scheduling complexity of  a connected communication graph of $V$, i.e.,  an assignment of power levels and colors to each link of the directed strongly connected graph such all transmissions are received correctly.

In this paper, we investigate the uniform connectivity problem, i.e., the connectivity problem for a set $V$ when only uniform power assignments are allowed. We give a formal definition of the graph we examine the connectivity of:

\begin{definition}[(Uniform) SINR graph]
Let $(M,d)$ be a metric space, $V \subseteq M$ be a finite set of \emph{nodes}, $c\,:\, V \rightarrow [k]$ be a \emph{coloring} of the nodes, and $E \subseteq V^2$ be the set defined as follows
\begin{equation}\label{eq:sinredge}
E = \left\{ (u,v) \in V^2 \;:\; \frac{1/d(u,v)^\alpha}{\sum_{w \in V \setminus \{ u \} \;:\; c(w) = c(u)} 1/d(w,v)^\alpha} \geqslant \beta \right\}
\end{equation}
We will refer to the directed graph $G  = (V,E)$ as the (uniform) \emph{SINR graph}.
\end{definition}

In other words, the definition of the graph says that a node $v$ can \emph{decode} a message coming from node $u$ (i.e. there is an edge from $u$ to $v$) if and only if the ratio between the \emph{power} (i.e. $1/d(u,v)^\alpha$) at which $v$ receives the message from $u$ and the sum of the powers from the other \emph{interfering} nodes (nodes $w$ that use the same \emph{frequency} or transmit in the same time slot, i.e. $c(w) = c(u)$) is at least some fixed constant $\beta$.

The question we want to answer is the following: Given the metric space $(M,d)$ and the set of nodes $V \subseteq M$, 
how many colors $k$ do we need in order to be sure that a coloring $c \,:\, V \rightarrow [k]$ exists such that the resulting graph $G$ is strongly connected?

In this paper, the set of nodes $V$ will be located in $\mathbb{R}$ or in $\mathbb{R}^2$ and $d$ will always denote the Euclidean distance.

\section{Connectivity in Grids}

\subsection{One-Dimensional Grid}
Let $V = \left\{ p_1, \dots, p_n \right\} \subseteq \mathbb{R}$ be a set of $n$ nodes with $p_1 < p_2 < \cdots < p_n$. We say that $V$ is a \emph{one-dimensional grid} if the nodes are equally spaced, i.e. $d(p_i,p_{i+1})$ is the same for every $i = 1, \dots, n-1$ (without loss of generality, we will assume $p_i = i$ for every $i$).

We say that a coloring $c \,:\, V \rightarrow [k]$ is a \emph{regular} $k$-coloring if the points are colored in a \emph{Round Robin} way, i.e. if $c(p_i) = (i \mod k) + 1$ for $i = 1, \dots, n$.

\begin{theorem}
Let $V = \{ p_1, \dots, p_n \}$ be a one-dimensional grid with $p_i = i$ for every $i=1, \dots, n$. For any $\alpha > 1$ a constant $k$ and a coloring $c \,:\, V  \rightarrow [k]$ exist such that the corresponding SINR graph is strongly connected.
\end{theorem}

\begin{proof}
Consider a regular $k$-coloring, where $k$ is a sufficiently large constant that we will choose later. Now we show that, for every $i=1,\dots, n-1$, there is a directed edge from node $p_i$ to node $p_{i+1}$ in the SINR graph. According to the definition of the SINR graph, we must show that
\begin{equation}\label{eq:constr}
\frac{1/d(p_i,p_{i+1})^\alpha}{\sum_{j \in [n] \setminus \{i\} \;:\; c(p_j) = c(p_i)} 1/d(p_j,p_{i+1})^\alpha} \geqslant \beta
\end{equation}
For the numerator, we have $1/d(p_i, p_{i+1})^\alpha = 1$ for any $\alpha$. For the denominator, observe that the nodes with the same color of $p_i$ are $\{ \dots p_{i-2k}, \; p_{i-k}, \; p_{i+k}, \; p_{i+2k}, \; \dots \}$. Thus, for any $j=1,\dots, n$, we have at most $2$ nodes at distance at least $j(k-1)$ from node $p_i$, hence
$$
\sum_{j \in [n] \setminus \{i\} \;:\; c(p_j) = c(p_i)} \frac{1}{d(p_j,p_{i+1})^\alpha}
\leqslant \sum_{j=1}^n \frac{2}{(j(k-1))^\alpha} = \frac{2}{(k-1)^\alpha}\sum_{j=1}^n \frac{1}{j^\alpha} < \frac{2}{(k-1)^\alpha} g(\alpha)
$$
where we named $g(\alpha) = \sum_{j=1}^\infty {j^{-\alpha}}$. Observe that $g(\alpha) = \mathcal{O}(1)$ for any constant $\alpha > 1$. In order to satisfy (\ref{eq:constr}) it is sufficient to choose $k \geqslant 1 + (2 \beta  g(\alpha))^{1/\alpha}$.

In exactly the same way we can show that for every $i=2,\dots, n$, there is a directed edge from node $p_i$ to node $p_{i-1}$, hence the the SINR graph is strongly connected.
\qed
\end{proof}

\subsection{Two-Dimensional Grid}

\begin{figure}[t]
\centering
\begin{tabular}{ccc}
\begin{minipage}{1.8in}
\centering
\includegraphics[width=1.8in]{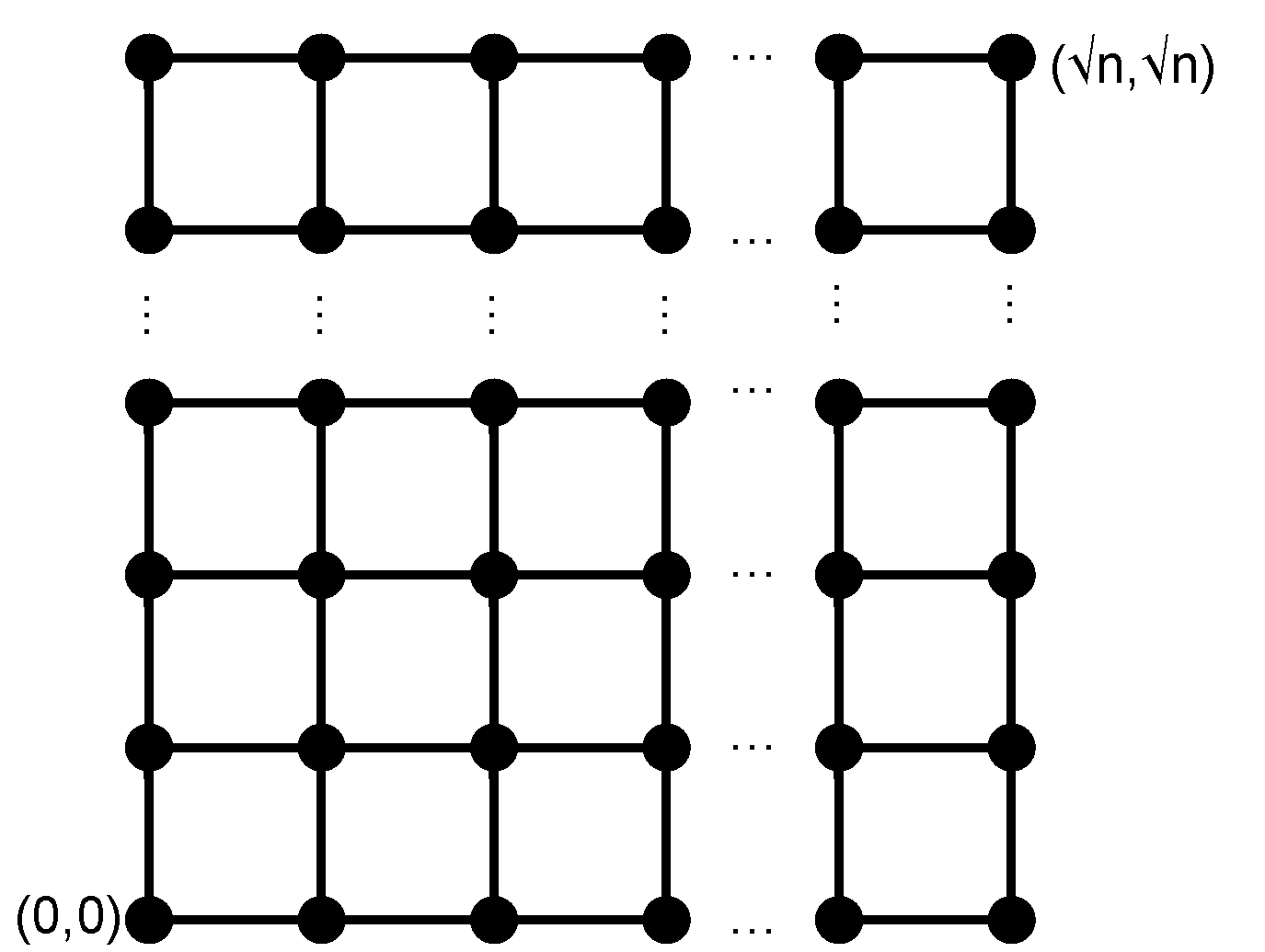}
\end{minipage}& &
\begin{minipage}{1.8in}
\centering
\includegraphics[width=1.8in]{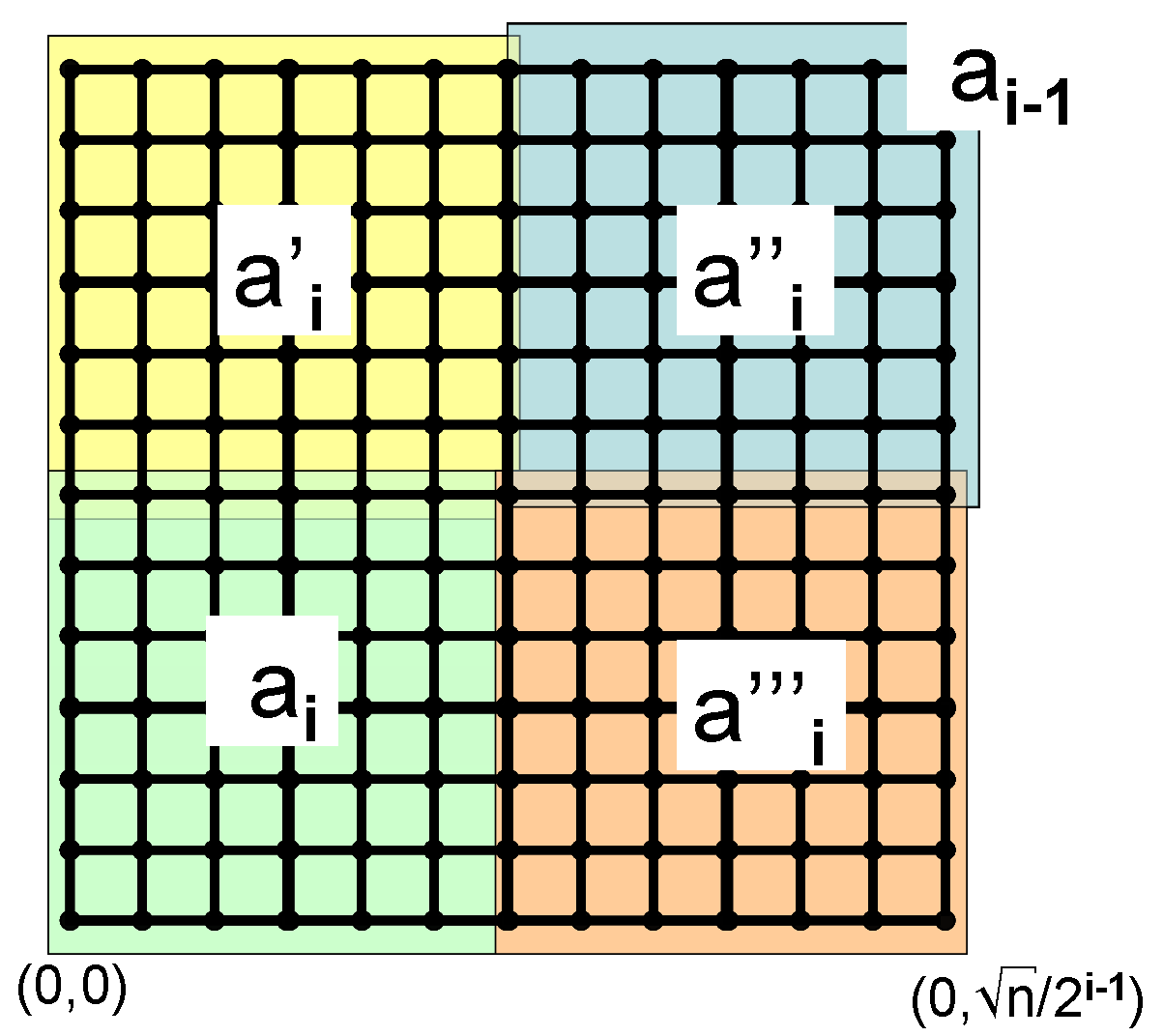}
\end{minipage}\\
\small (a) & & \small (b)\\
\end{tabular}
\caption{(a) Two dimensional grid topology. (b) Grid division for the lower bound.}
\label{fig:grid}
\end{figure}

Consider the following two dimensional grid topology of $n$ nodes. An array of $\sqrt{n}$ arrays containing $\sqrt{n}$ nodes each, where the left bottom corner node is denoted by (0,0), see Fig. \ref{fig:grid} (a).

A \emph{regular $k^2$-coloring} partitions the nodes into $k^2$ sets such that the closest distance between any two nodes of the same color is $k$. In other words, each set forms another grid with distance $k$. If $\alpha$ exceeds two, the number of colors required for connectivity is constant.

\begin{theorem}[Bound 2D grids, $\alpha>2$]\label{thm:b2Dgridalpha>2}
Let $V = \{ p_1, \dots, p_{n} \} \subseteq [0,\sqrt{n}]^2$ be a two-dimensional grid. For any $\alpha > 2$ a constant $k$ and a coloring $c \,:\, V  \rightarrow [k]$ exist such that the corresponding SINR graph is strongly connected.
\end{theorem}
\begin{proof}
Consider a regular $k^2$-coloring for a grid consisting of $n$ nodes. Let the node $v$ at (0,0) belong to color $j$. Without loss of generality we can assume that $v$ is connected to the node at (0,1) in the corresponding interference graph. We now explore the interference accrued at node (0,1) if all nodes of color $j$ transmit simultaneously. In this case the total interference at (0,1) is
$$
I_{(0,1)}<\sum_{i=1}^{\sqrt{n}}{\frac{2i+1}{(ki-1)^\alpha}}<\frac{3}{(k/2)^\alpha} \sum_{i=1}^{\sqrt{n}}{\frac{1}{i^{\alpha-1}}}<\frac{3\cdot2^\alpha(\alpha-1)}{2k^\alpha(\alpha-2)},
$$
for $\alpha>2$ due to a standard bound for Rieman's zeta-function. This level of interference needs to be below $1/\beta$, hence the distance $k$ has to satisfy the following inequality: $k>\left(3\cdot2^{\alpha-1}\beta(\alpha-1)/(\alpha-2)\right)^{1/\alpha}.$

Note that the node in the center of the grid faces at most four times the amount of interference that the node at (0,1) is exposed to. Thus this procedure can be repeated to bound the interference at any node in the grid.  In other words, a regular $\left(6\cdot2^\alpha\beta(\alpha-1)/(\alpha-2)\right)^{2/\alpha}$-coloring ensures connectivity in a constant number of rounds.\qed
\end{proof}

Observe that this result holds for infinite grids as well. In addition, it coincides with the UDG interference model, where a constant number of colors suffices as well. The situation changes dramatically if $\alpha$ is less than or equal to two. If $\alpha=2$, the number of necessary colors increases logarithmically in the number of nodes.

\begin{theorem}[Upper bound 2D grids, $\alpha=2$]\label{thm:ub2Dgrid}
Let $V = \{ p_1, \dots, p_n \} \subseteq [0,\sqrt{n}]^2$ be a two-dimensional grid. For $\alpha = 2$ a regular $\mathcal{O}(\log n)$-coloring ensures that the corresponding SINR graph is strongly connected.
\end{theorem}

\begin{proof}
We start similarly to the proof for $\alpha>2$ and sum up the interference accumulated at node (0,1) under a regular $k^2$-coloring. In this case the total interference at (0,1) is less than
$$
I_{(0,1)}<\sum_{i=1}^{\sqrt{n}}{\frac{2i+1}{(ki-1)^\alpha}}<\frac{3}{(k/2)^\alpha} \sum_{i=1}^{\sqrt{n}}{\frac{1}{i^{\alpha-1}}}=\frac{6\log{n}}{k^2}.
$$
Moreover, the total interference at (0,1) exceeds
$$
I_{(0,1)}>\sum_{i=1}^{\sqrt{n}}{\frac{2i+1}{(\sqrt{2}ki)^\alpha}}>\frac{\sqrt{2}^\alpha}{d^\alpha} \sum_{i=1}^{\sqrt{n}}{\frac{1}{i^{\alpha-1}}}=\frac{2\log{n}}{k^2}.
$$
Note that the node in the center of the grid faces at most four times the amount of interference that the node at (0,1) is exposed to.

$\beta$ being a constant entails that $k^2$ has to be in the order of $\Omega(\log{n})$ if we want that a message from the node at (0,0) can be decoded at (0,1). There are $\mathcal{O}(k^2)$ nodes at a radius of $k$ around (0,1), consequently, we need $\Omega(\log{n})$ frequencies if $\alpha=2$ and we want all nodes to be able to send concurrently and form a connected structure. We achieve this goal by partitioning the existing grid into $\log{n}$ grids that send with distinct frequencies. \qed
\end{proof}

\begin{theorem}[Lower bound 2D grids, $\alpha=2$]\label{thm:lb2Dgrid}
Let $V = \{ p_1, \dots, p_n \} \subseteq [0,\sqrt{n}]^2$ be a two-dimensional grid and $\alpha = 2$.
Let $c \,:\, V  \rightarrow [k]$ be a coloring. If the corresponding SINR graph is strongly connected, then the number of colors is $k = \Omega\left(\frac{\log n}{\log \log n}\right)$.
\end{theorem}

\begin{proof}
For the lower bound, we show that no matter how we distribute the colors on the grid, we need $\Omega(\frac{\log n}{\log \log n})$ colors to ensure connectivity. More precisely, we show that in whatever way we position the nodes, we can always find a node where the interference experienced is at least as high as at (0,0) in the grid situation.

Let us start by demonstrating the minimum interference accumulated at any node if we use three colors. Without loss of generality, there is at least one color $j$ that is assigned to at least $\frac{n}{3}$ nodes. In the following, we will only consider this color $j$. Let us divide the grid into 4 parts $(a_1, a_1', a_1'', a_1''')$ of equal size. Among these, there is at least one square with at least $n/12$ nodes with color $j$, because there would not be $\frac{n}{3}$ nodes of color $j$ together with the other squares otherwise. Without loss of generality we can assume that this is the square $a_1$ anchored in (0,0) and we denote the number of nodes in $a_1$ by $|a_1|$. We now want to compute the minimal interference that one of the nodes in $a_1$ experiences. To this end we assume that there are exactly $\frac{n}{3}$ nodes with color $j$ and exactly $\frac{n}{12}$ nodes in $a_1$ (otherwise the interference for nodes in $a_1$ increases. By positioning all $\frac{n}{3}-|a_1|=\frac{n}{4}$ nodes that are not in $a_1$ into the corner $(\sqrt{n},\sqrt{n})$, i.e. the corner with the largest distance from (0,0), the minimal interference any node in $a_1$ experiences exceeds $\frac{n}{4}\cdot\frac{1}{2n}=\frac{1}{8}$ because the largest distance between a point in $a_1$ and $(\sqrt{n},\sqrt{n})$ is $\sqrt{2n}$. Let us now consider the interference the nodes in $a_1$ cause among themselves. We proceed as before by dividing the square $a_1$ into four squares $(a_2, a_2', a_2'', a_2''')$ of side length $\frac{\sqrt{n}}{4}$ each. Using the same arguments we know that one of them, let us say, $a_2$ contains at least $\frac{n}{48}$ nodes  of color $j$ and to minimize the interference within $a_2$ we look at the case where $|a_2|$ is $\frac{n}{48}$ and anchored at (0,0). We can now compute the minimal amount of interference caused by the $\frac{n}{12}-\frac{n}{48}=\frac{n}{16}$ nodes in $a_1$ at (0,0) to be at least $\frac{n}{16}\cdot\frac{2}{n}=1/8$ because the maximal distance within $a_1$ is $\frac{\sqrt{n}}{\sqrt{2}}$. If we repeat these steps, it holds that in step $i$ we have $\frac{n}{4^i}$ nodes in distance $\frac{\sqrt{2n}}{2^{i-1}}$ responsible for a sum of interference of $\frac{n}{4^i}\cdot \frac{4^{i-1}}{2n}=1/8$ (see Fig. \ref{fig:grid} (b)). After $\lfloor \log_4 n \rfloor$ steps there is only one node left in $a_i$ and we stop. The total interference is thus in $\Omega(\log n)$.

We can generalize this approach to more than three colors. If we use $k$ colors and partition the square with most nodes into $k+1$ squares and proceed recursively, the number of nodes in $a_{i-1}$ outside $a_i$ is in the order of $\frac{n}{(k+1)^i}$, where $a_i$ is the square with most nodes in the $i^{th}$ step. These nodes are at most in distance $\frac{\sqrt{2n}}{(k+1)^{(i-1)/2}}$ from the nodes in $a_i$ and thus cause interference of $\frac{n}{(k+1)^i}\cdot\frac{(k+1)^{i-1}}{2n}=\frac{1}{2(k+1)}$. The maximal number of recursions is $\lfloor \frac{\log n}{\log k} \rfloor$. Consequently, all the nodes are responsible for $\Omega(\frac{\log n}{k \log k})$ interference at (0,0).

Typically the SINR threshold $\beta$ that guarantees the reception of a message is a small constant. Hence, a neighbor on the grid (at distance 1) can receive our message if and only if the total interference is at most $1/\beta$. Now observe that, if $k < \frac{\log n}{c \log \log n}$ for some positive constant $c$, then we have that
$$
\frac{\log n}{k \log k} > \frac{\log n}{\frac{\log n}{c \log \log n}(\log \log n - \log (c\log\log n))} > c
$$
Thus, for any constant $\beta$, a large enough constant $c$ exists such that, if $k < \frac{\log n}{c \log \log n}$, then the interference ratio is larger than $\beta$.
\qed
\end{proof}

\begin{corollary}[Upper and lower bound 2D grids, $1 \leqslant \alpha < 2$]
Let $V = \{ p_1, \dots, p_n \} \subseteq [0,\sqrt{n}]^2$ be a two-dimensional grid and $1 \leqslant \alpha < 2$.
Let $c \,:\, V  \rightarrow [k]$ be a coloring. If the corresponding SINR graph is strongly connected, then the number of colors is $k = \Theta\left(n^{2/\alpha-1}\right)$.
\end{corollary}

\begin{proof}
Given a regular $k^2$-coloring, the total interference at any point in [0,$\sqrt{n}^2$] is less than
$$I<4\sum_{i=1}^{\sqrt{n}}{\frac{2i+1}{(ki-1)^\alpha}}<\frac{3\cdot 2^{2+\alpha}}{(k/2)^\alpha} \sum_{i=1}^{\sqrt{n}}{\frac{1}{i^{\alpha-1}}}<\frac{3\cdot 2^{2+\alpha}n^{1-\alpha/2}}{(2-\alpha)k^\alpha}.$$
Due to the same arguments as in the proof of Theorem \ref{thm:ub2Dgrid}, this implies that a regular coloring using $\mathcal{O}(n^{2/\alpha-1})$ colors suffices for connectivity.

For the matching lower bound we adopt the same recursive strategy as in the proof of Theorem \ref{thm:lb2Dgrid}. In step $i$ we have $\frac{n}{(k+1)^i}$ nodes in distance $\frac{\sqrt{2n}}{(k+1)^{(i-1)/2}}$ responsible for a sum of interference of
$$\frac{n}{(k+1)^i}\cdot \frac{(k+1)^{\alpha(i-1)/2}}{(2n)^{\alpha/2}}=\frac{n^{1-\alpha/2}}{2^{\alpha/2}(k+1)^{\alpha/2}}(k+1)^{i(\alpha/2-1)}.$$
After $\lfloor \log_{k+1} n \rfloor$ steps there is only one node left in $a_i$ and we stop. The total interference is thus
\begin{eqnarray*}
I &= &\frac{n^{1-\alpha/2}}{(2(k+1))^{\alpha/2}}\sum_{i=1}^{\log_{k+1} n}(k+1)^{i(\alpha/2-1)}\\
&=&\frac{n^{1-\alpha/2}}{(2(k+1))^{\alpha/2}} \cdot c,
\end{eqnarray*}
for some constant $c$, since $(k+1)^{\alpha/2-1}<1$. Hence, in order to make sure that a message in distance one from a sender can be received, i.e. $1/I>\beta$, the smallest possible $k$ has to be in the order of $\Omega(n^{2/\alpha-1})$.
\qed
\end{proof}

\section{Connectivity for Random Instances: The One-Dimensional Case}
In this section, we consider a set $V$ of $n$ nodes thrown independently and uniformly at random in $[0,1]$, the unit interval.\footnote{In contrast to the grid, where we set the minimal distance between two nodes to be one, we consider the unit interval for the random case because of its direct correspondence to probability.} We assume the path-loss exponent to be $\alpha = 2$.

Our first result shows that $\mathcal{O}(\log n)$ colors are enough to guarantee strong connectivity of the corresponding SINR graph.

\begin{theorem}[Upper bound]\label{theorem:rrub}
Let $V = \{ p_1, \dots, p_n \} \subseteq [0,1]$ where $p_1, \dots p_n$ are independent random variables uniformly distributed in $[0,1]$. Then a coloring $c \,:\, V  \rightarrow [k]$ exists, with number of colors $k = \mathcal{O}(\log n)$, such that the corresponding SINR graph is strongly connected w.h.p.\footnote{We use the shortcut \emph{w.h.p.} (\emph{with high probability}) to say that the event holds with probability at least $1-n^{-c}$, for some positive constant $c$}
\end{theorem}

\begin{quote}
\ideaproof
Consider a \emph{regular} coloring of $c \log n$ colors, with a sufficiently large constant $c$, so we can partition the interval $[0,1]$ in subintervals of length $\Theta(\log n / n)$, each one of them containing (i) $\Theta(\log n)$ nodes w.h.p. and (ii) no more than one node for each color w.h.p.

\noindent
For any node $p$, we can take an interval of length $\Theta(\log n / n)$ containing $\Theta(\log n)$ nodes and such that every node in that interval is an out-neighbor of node $p$. Indeed, for any node $q$ in that interval, the power at which $q$ receives the signal from $p$ is $\Omega(n^2 / \log^2 n)$. For the nodes \emph{interfering} with $p$, we have that for any $h$ there are $\mathcal{O}(1)$ interfering nodes at distance $\Omega(h \log n / n)$ from $q$, hence the total \emph{interfering power} at node $q$ is $\mathcal{O}(n^2 / \log^2 n)$. By choosing the constant $c$ appropriately, the resulting ratio between the power at which $q$ receives the signal from $p$ and the interfering power is an arbitrary large constant.
\qed
\end{quote}
\begin{proof}
Let $I \subseteq [0,1]$ be an interval of length $c \log n / n$, where $c$ is a sufficiently large constant that we will choose later, and let $X$ be the r.v. counting the number of nodes in $I$. The expectation of $X$ is $\Expec{X} = c \log n$. Since $X$ can be written as sum of independent Bernoulli r.v., we can use the Chernoff bound (\ref{eq:cbr}) with $\delta = 1/2$ and $\mu = c \log n$ to obtain
\begin{equation}\label{eq:nii}
\Prob{X > \frac{3c}{2} \log n} < n^{-c/12}.
\end{equation}
Set the number of colors to be $k = \frac{3c}{2} \log n$, and consider a \emph{regular} coloring, i.e. the color of node $p_i$ is $(i \mod k) + 1$ for $i = 1, \dots, n$.

For a node $p_i$ let $I_i$ be an interval of length $32 \log n / n$ centered in $p_i$ (shorter if $p_i$ is close to the boundary).
$$
I_i = \left[ p_i - 16 \frac{\log n}{n}, \; p_i + 16 \frac{\log n}{n} \right] \cap [0, \, 1].
$$
Let $Y_i$ be the number of nodes in $I_i$ and observe that $\Expec{Y_i} \geqslant 16 \log n$. By using Chernoff bound (\ref{eq:cbl}) with $\delta = 1/2$ and $\mu = 16 \log n$, we have that
$$
\Prob{Y_i < 8 \log n} < n^{-2}.
$$
Now we show that every node in interval $I_i$ is w.h.p. an \emph{out-neighbor} of node $p_i$ in the SINR graph. This will prove that the SINR graph is strongly connected w.h.p. Consider the interval
$$
J_i = \left[ p_i - \frac{c}{2} \frac{\log n}{n}, \; p_i + \frac{c}{2} \frac{\log n}{n} \right] \cap [0, \, 1].
$$
and partition the rest of $[0,1]$ into intervals of length $c \log n / n$ (possibly shorter for the two intervals on the boundary). Now observe that it follows, from (\ref{eq:nii}) and the fact that we are using a $k$-regular coloring, that the probability of a single interval containing more than one node with the same color is less than $n^{-c/12}$. Due to the union bound over all intervals, the probability that one interval exists that contains more than one color is less than $n^{-c/12 + 1}$. Now, conditioning on the event that all such intervals contain no more than one node with the same color, we can show that every node in $I_i$ is an out-neighbor of node $p_i$.

\noindent Let $q \in I_i$ be a node, such that we have for the numerator in (\ref{eq:constr})
$$
\frac{1}{d(q, p_i)^2} \geqslant \frac{n^2}{16^2 \log^2 n}.
$$
For the denominator, observe that, for any $h=1, \dots, n$, we have at most two nodes with the same color as $p_i$, at distance at least $h (c/2 - 16) \log n / n \geqslant h (c/4) \log n / n$, where we assume $c \geqslant 64$ in the inequality. Hence,
$$
\sum_{j \in [n] \setminus \{i\} \;:\; c(p_j) = c(p_i)} \frac{1}{d(p_j,q)^2} \leqslant
\sum_{h=1}^n \frac{2}{\left( h \frac{c}{4} \frac{\log n}{n} \right)^2} = \frac{32}{c^2} \frac{n^2}{\log^2 n} \sum_{h=1}^n \frac{1}{h^2} \leqslant \frac{16 \pi^2}{3c^2} \frac{n^2}{\log^2 n}.
$$
In order to satisfy (\ref{eq:constr}), it is sufficient to choose $c > 64 \pi \sqrt{\beta / 3}$, and it follows that
$$
\frac{1/d(p_i,q)^2}{\sum_{j \in [n] \setminus \{i\} \;:\; c(p_j) = c(p_i)} 1/d(p_j,q)^2} \geqslant \frac{\frac{n^2}{16^2 \log^2 n}}{\frac{16 \pi^2}{3c^2} \frac{n^2}{\log^2 n}} = \frac{3}{16^3 \pi^2} c^2 \geqslant \beta.
$$
Hence, the probability that node $p_i$ is not connected to all nodes in $I_i$, or $I_i$ does not contain $c/8 \log n$ nodes is less than $n^{-2} + n^{-c/12}$. Due to the union bound on all nodes we can deduce the probability that a node exists that is not connected to at least $\Omega(\log n)$ nodes in an interval of length $\mathcal{O}(\log n / n)$ to be $\mathcal{O}(1/n)$. The SINR graph is thus connected w.h.p. \qed
\end{proof}

\medskip\noindent
The previous theorem shows that, with a regular $\mathcal{O}(\log n)$-coloring, the resulting SINR graph is strongly connected w.h.p. Now we prove that this is the best we can achieve with regular colorings.

The next lemma provides a condition implying that the SINR graph is not strongly connected. We will use it in the proof of Theorem~\ref{prop:rrlb}.

\begin{lemma}\label{lemma:lb}
Let $V = \{ p_1, \dots, p_n \} \subseteq [0,1]$ be an arbitrary set of nodes and let $c \,:\, V  \rightarrow [k]$ be a regular coloring. Suppose that a length $0 < \ell < 1/3$ and a point $x \in [0, \, 1 - 3 \ell]$ exist such that the following conditions hold:
\begin{enumerate}
\item In the subinterval $[x, \, x + \ell]$, there are at least $(4/\beta)k$ nodes;
\item In the subinterval $[x + \ell, \, x + 2 \ell]$, there are no nodes;
\item In the subinterval $[x + 2\ell, \, x + 3 \ell]$ there is at least one node.
\end{enumerate}
Then, the SINR graph is not strongly connected.
\end{lemma}
\begin{proof}
Let $p$ be a node in $[0, x + \ell]$ and let $q$ be a node in $[x + 2 \ell, 1]$. Due to hypothesis (1) and the fact that the coloring is regular, there are at least $(4/\beta)$ nodes in the interval $[x, \, x + \ell]$ interfering with the transmission from $p$ to $q$, that are at distance less than $\ell + d(p,q)$ from $q$. Hence, the interference ratio at node $q$ is less than
$$
\frac{ \frac{1}{d(p,q)^2}}{\frac{4/\beta}{\left( d(p,q) + \ell \right)^2} }
= \frac{ \frac{1}{d(p,q)^2}}{\frac{4/\beta}{d(p,q)^2 \left( 1 + \frac{\ell}{d(p,q)} \right)^2} }
= \frac{ \beta \left( 1 + \frac{\ell}{d(p,q)} \right)^2}{4} < \beta.
$$
In the last inequality we used $d(p,q) > \ell$. Thus, there are no edges from nodes in $[0, \, x + \ell]$ to nodes in $[x + 2 \ell, \, 1 ]$ and by hypothesis (2) the graph is not strongly connected.
\qed
\end{proof}

\begin{theorem}[Lower Bound for regular colorings]\label{prop:rrlb}
Let $V = \{ p_1, \dots, p_n \} \subseteq [0,1]$ where $p_1, \dots p_n$ are independent random variables uniformly distributed in $[0,1]$, and let $c \,:\, V  \rightarrow [k]$ be a regular coloring. If the corresponding SINR graph is strongly connected w.h.p., then the number of colors is $k = \Omega(\log n)$.
\end{theorem}
\begin{proof}
Let $\ell = (4/\beta)(k/n)$ and let $I \subseteq [0,1]$ be an interval of length $3\ell$. Consider the event
$$
\mathcal{E}_I = \mbox{``Interval $I$ satisfies conditions (1), (2), and (3) of Lemma~\ref{lemma:lb}''}
$$
In what follows we prove that, if $k < (\beta/2) \log n$, then an interval $I$ of length $3 \ell$ exists such that $\mathcal{E}_I$ holds w.h.p. To this end we use the Poisson approximation (for a detailed description of this approach see, for example, Chapter 5.4 in~\cite{MU05}).

For $i = 1, \dots, n$ let $X_i$ be the random variable counting the number of nodes in the interval $[ (i-1)/n, \, i/n ]$ and observe that $\Expec{X_i} = 1$. Consider the set $\left\{ I_0, \dots, I_h \right\}$ of disjoint intervals of length $3\ell$, where $I_j = [3 \ell j , \, 3 \ell(j+1) ]$ for $j=0,1, \dots, h$ and observe that, since $\ell = \mathcal{O}(\log n / n)$, then the number of such intervals is $h = \Omega(n/\log n)$. For interval $I_j$ we can write the event $\mathcal{E}_{I_j}$ as
$$
\mathcal{E}_{I_j} = \left( \sum_{i=3 n \ell j + 1}^{3 n \ell j + n \ell} X_i \geqslant n \ell \right)
\cap \left( \sum_{i=3 n \ell j + n \ell + 1}^{3 n \ell j + 2 n \ell} X_i = 0 \right)
\cap \left( \sum_{i=3 n \ell j + 2 n \ell + 1}^{3 n \ell j + 3 n \ell} X_i \geqslant 1 \right).
$$
Now let $Y_1, \dots, Y_n$ be i.i.d. Poisson random variables with $\Expec{Y_i}=1$ and let $\mathcal{F}_{I_j}$, for $j=0,1, \dots, h$, be the events in the Poisson setting corresponding to the events $\mathcal{E}_{I_j}$, i.e.
$$
\mathcal{F}_{I_j} = \left( \sum_{i=3 n \ell j + 1}^{3 \ell j + n \ell} Y_i \geqslant n \ell \right)
\cap \left( \sum_{i=3 n \ell j + n \ell + 1}^{3 n \ell j + 2 n \ell} Y_i = 0 \right)
\cap \left( \sum_{i=3 n \ell j + 2 n \ell + 1}^{3 n \ell j + 3 n \ell} Y_i \geqslant 1 \right).
$$
Since the $Y_i$s are independent, it holds that
\begin{eqnarray*}
\Prob{\mathcal{F}_{I_j}} & = & \Prob{\left( \sum_{i=3 n \ell j + 1}^{3 n \ell j + n \ell} Y_i \geqslant n \ell \right)
\cap \left( \sum_{i=3 n \ell j + n \ell + 1}^{3 n \ell j + 2 n \ell} Y_i = 0 \right)
\cap \left( \sum_{i=3 n \ell j + 2 n \ell + 1}^{3 n \ell j + 3 n \ell} Y_i \geqslant 1 \right)} \\
& = & \Prob{\sum_{i=3 n \ell j + 1}^{3 n \ell j + n \ell} Y_i \geqslant n \ell}
\cdot \Prob{\sum_{i=3 n \ell j + n \ell + 1}^{3 n \ell j + 2 n \ell} Y_i = 0}
\cdot \Prob{\sum_{i=3 n \ell j + 2 n \ell + 1}^{3 n \ell j + 3 n \ell} Y_i \geqslant 1} \\
& \geqslant &\frac{1}{e} \cdot e^{-n \ell} \cdot \left( 1 - e^{-n \ell} \right) \geqslant \frac{1}{2e} e^{-n \ell} = \frac{1}{2e} e^{-(4/\beta)k}.
\end{eqnarray*}
Thus, if $k < (\beta/2) \log n$ then $\Prob{\mathcal{F}_{I_j}} \geqslant \frac{1}{2e \sqrt{n}}$ and, since the intervals $I_j$ are disjoint, the probability that no one of the events $\mathcal{F}_{I_j}$ happens is
$$
\Prob{\bigcap_{j=0}^{h} \overline{\mathcal{F}_{I_j}} } = \prod_{j=0}^{h} \Prob{\overline{\mathcal{F}_{I_j}}} \leqslant \left( 1 - \frac{1}{2e \sqrt{n}} \right)^{h} \leqslant e^{-\frac{h}{2e\sqrt{n}}}.
$$
The Poisson approximation implies that the probability that none of the events $\mathcal{E}_{I_j}$ occurs is
$$
\Prob{\bigcap_{j=0}^{h} \overline{\mathcal{E}_{I_j}}} \leqslant e \sqrt{n} \cdot \Prob{\bigcap_{j=0}^{h} \overline{\mathcal{F}_{I_j}} } \leqslant e \sqrt{n} e^{-\frac{h}{2e\sqrt{n}}}.
$$
Since $h = \Omega(n / \log n)$ this probability is exponentially small. Hence, at least one of the intervals $I_j$ satisfies conditions (1), (2), and (3) of Lemma~\ref{lemma:lb} w.h.p.
\qed
\end{proof}

\subsection{Lower bound for arbitrary colorings}

\smallskip\noindent
In Theorem~\ref{theorem:rrub} we showed that, using a regular coloring with $\mathcal{O}(\log n)$ colors, we can make the SINR graph strongly connected. In Therorem~\ref{prop:rrlb} we proved that, if we restrict ourselves to regular colorings, we cannot use asymptotically less colors. An interesting open question is whether or not we can find a \emph{non-regular} coloring with $o(\log n)$ colors that makes the SINR strongly connected. In what follows we prove that, in any case, we must use at least $\Omega(\log \log n)$ colors.

\begin{definition}[Exponential sequence]
Let $V = \{ q_1, \dots, q_h \} \subseteq [a,b] \subseteq [0,1]$ be a set of \emph{nodes} in a subinterval $[a,b]$ of the unit interval and let $n \geqslant 2^h$. We say that $V$ is an \emph{exponential sequence} if a constant $0 < \varepsilon < 1/3$ exists such that $(1-\varepsilon) 2^i/n \leqslant q_i - a \leqslant (1+\varepsilon) 2^i/n$ for every $i=1, \dots, h$.
\end{definition}

\noindent
Observe that if $V$ is an exponential sequence, than for every $i$ it holds that
$$
(1 - 3 \varepsilon) 2^i/n \leqslant d(q_i, q_{i+1}) \leqslant (1 + 3 \varepsilon) 2^i/n
$$

\begin{lemma}\label{lemma:lbloglog}
Let $V = \{ p_1, \dots, p_n \} \subseteq [0,1]$ be a set of nodes and let $c\,:\, V \rightarrow [k]$ be a colouring of $V$ with $k$ colors. If an interval $[a,b] \subseteq [0,1]$ exists such that $V \cap [a,b] = \{ q_1, \dots, q_h \}$ is an exponential sequence with $h$ nodes and if $k < \gamma h$ where
\begin{equation}\label{eq:gammacond}
\gamma = \frac{\beta}{\beta + \left( 1 + \frac{1-\varepsilon}{1 - 3 \varepsilon} \right)^\alpha}
\end{equation}
then the corresponding SINR graph is not strongly connected. 
\end{lemma}
\begin{proof}
If the number of colours is $k < \gamma h$, then there are at least $1/\gamma$ nodes in the exponential sequence $\{ q_1, \dots, q_h \}$ with the same colour $j \in [k]$. Let $V_j \subseteq V$ be the set of nodes with colour $j$, and let $q_i$ be the rightmost node with such color, $q_i = \max V_j$. Now we show that, in the SINR graph, node $q_i$ has no out-neighbours.

Let $x \in V$ be a node with $x > q_i$, let us name $\ell$ the distance between $x$ and $q_i$ and observe that $\ell = d(q_i,x) \geqslant d(q_i,q_{i+1}) \geqslant (1 - 3 \varepsilon)2^i/n$. In the interval $[a,b]$ there are at least $1/\gamma - 1$ nodes $w \in V_j$ with the same colour of $q_i$, and for each one of them, the distance from $x$ is
$$
d(w,x) = d(w,q_i) + \ell \leqslant (1 + \varepsilon) 2^i/n + \ell
$$
Hence, the interference ratio is
$$
\frac{\frac{1}{d(q_i,x)^\alpha}}{\sum_{w \in V_j} \frac{1}{d(w,x)^\alpha}}
< \frac{ \frac{1}{\ell^\alpha} }{ \frac{1/\gamma - 1}{\left((1 + \varepsilon) 2^i/n + \ell \right)^\alpha}}
= \frac{ \frac{1}{\ell^\alpha} }{ \frac{1/\gamma - 1}{\ell^\alpha \left(1 + \frac{(1 + \varepsilon) 2^i/n}{\ell} \right)^\alpha}}
= \frac{ \left(1 + \frac{(1 + \varepsilon) 2^i/n}{\ell} \right)^\alpha }{ 1/\gamma - 1}
\leqslant \frac{ \left(1 + \frac{1 + \varepsilon}{1 - 3 \varepsilon} \right)^\alpha }{ 1/\gamma - 1}
$$
From (\ref{eq:gammacond}), it follows that the ratio is smaller than $\beta$. Hence, node $q_i$ has no out-neighbours \emph{on the right}.

Exactly in the same way, it is easy to see that node $q_i$ has no out-neighbours \emph{on the left}, thus the SINR graph is not strongly connected.
\qed
\end{proof}

\begin{theorem}[Lower Bound for arbitrary colorings]\label{theorem:arlb}
Let $V = \{ p_1, \dots, p_n \} \subseteq [0,1]$ where $p_1, \dots p_n$ are independent random variables uniformly distributed in $[0,1]$, and let $c \,:\, V  \rightarrow [k]$ be any coloring. If the corresponding SINR graph is strongly connected w.h.p., then the number of colors is $k = \Omega(\log \log n)$.
\end{theorem}
\begin{quote}
\ideaproof
Split the interval $[0,1]$ in $\Theta(n / \log n)$ disjoint intervals of length $\Theta(\log n / n)$. Choose the constants in a way that, for at least one of such intervals $I$, it holds that $V \cap I$ is an exponential sequence of length $\Theta(\log \log n)$ w.h.p.
From Lemma~\ref{lemma:lbloglog} it follows that $\Omega(\log \log n)$ colors are needed in order to have a strongly connected SINR graph.
\qed
\end{quote}
\begin{proof}
Let $I = \left[ \frac{a}{n}, \, \frac{b}{n} \right] \subseteq [0,1]$ be an interval with $b-a = \ell = \left\lfloor (1/2) \log n \right\rfloor$. Consider the event
$$
\mathcal{E}_I = \mbox{``} V \cap I \mbox{ is an exponential sequence of length } \Omega(\log \log n) \mbox{''}
$$
For $i = 1, \dots, \ell$ let $X_i$ be the random variable counting the number of nodes in the interval
$$
\left[ \frac{a+i-1}{n}, \, \frac{a+i}{n} \right]
$$
and observe that $\Expec{X_i} = 1$ for every $i$. Consider the set of indices
$$
J = \left\{ 2^j \,:\, j = 1, \dots, \lfloor \log \log n \rfloor - 1 \right\} \subseteq \left\{ 1, 2, \dots, \ell \right\}
$$
If $X_i = 1$ for every $i \in J$ and $X_i = 0$ for every $i \in [\ell] \setminus J$ then the event $\mathcal{E}_I$ holds. Hence
$$
\Prob{\mathcal{E}_I} \geqslant \Prob{ \left( \bigcap_{i \in J} \{ X_i = 1 \} \right) \cap \left( \bigcap_{i \in [\ell] \setminus J} \{ X_i = 0 \} \right) }
$$

\noindent
Let $Y_1, \dots, Y_\ell$ be i.i.d. Poisson random variables with expectation $1$, and let $\mathcal{F}_I$ be the event corresponding to $\mathcal{E}_I$ in the Poisson setting. Then
\begin{eqnarray*}
\Prob{\mathcal{F}_I} & \geqslant & \Prob{ \left( \bigcap_{i \in J} \{ Y_i = 1 \} \right) \cap \left( \bigcap_{i \in [\ell] \setminus J} \{ Y_i = 0 \} \right) } \\
& = & \prod_{i \in J} \Prob{ Y_i = 1 } \prod_{i \in [\ell] \setminus J} \Prob{ Y_i = 0 } = e^{-\ell} \geqslant \frac{1}{\sqrt{n}}
\end{eqnarray*}

\noindent
Now consider a set of $\Omega(n / \log n)$ disjoint intervals $I_1, \dots, I_h$, each one of length $\ell = \left\lfloor (1/2) \log n \right\rfloor$. Then, by using the Poisson approximation (for a detailed description of such tool see, for example, Chapter 5.4 in~\cite{MU05}), the probability that no one of them gives an exponential sequence is
$$
\Prob{\bigcap_{j=1}^h \overline{\mathcal{E}_{I_j}}}  \leqslant e \sqrt{n} \Prob{\bigcap_{j=1}^h \overline{\mathcal{F}_{I_j}}} = e \sqrt{n} \prod_{j=1}^h \Prob{\overline{\mathcal{F}_{I_j}}} \leqslant e \sqrt{n} \left( 1 - \frac{1}{\sqrt{n}} \right)^h \leqslant e \sqrt{n} e^{- h/\sqrt{n}}
$$
Since $h = \Omega(n / \log n)$, the above probability is exponentially small. Thus, for at least one of the intervals it holds that $V \cap I_j$ is an exponential sequence w.h.p.
\qed
\end{proof}

\section{Conclusions and open problems}
In this paper we initiate the study of connectivity in the uniform power SINR model. Clearly we can not achieve connectivity in the SINR model if we use only one frequency, since the SINR diagram is a partition of the plane. To overcome this problem we can either use a sophisticated scheduling algorithm or we can increase the number of frequencies. However those two actions are equivalent i.e., any schedule can be translated into a choice of frequencies and any frequency assignment can be translated into a schedule. Therefore we can defined the connectivity problem in the SINR model as the minimal number of frequency the network needs to use to maintain connectivity (the scheduling complexity of connectivity).

We provided upper and lower bounds for the number of time slots or frequencies to build a strongly connected graph of communication edges. We focused on nodes arranged in a regular grid or uniformly at random on the unit interval.
We proved that if the nodes are located on a regular grid the number of frequencies needed to maintain connectivity is a function of the dimension of the grid and the path-loss exponent $\alpha$. Apart from the special case $\alpha=2$ these bounds are asymptotically tight. In contrast, when transmitters are located uniformly at random on the interval $[0,1]$ there is a big gap between the upper bound $\mathcal{O}(\log n)$ in Theorem~\ref{theorem:rrub} and the lower bound $\Omega(\log \log n)$ in Theorem~\ref{theorem:arlb}. A natural open question is to close this gap. Other intriguing problems include determining upper and lower bounds for general colorings in the random two-dimensional case, or algorithms computing the uniform power complexity of connectivity of arbitrarily positioned nodes.

\begin{tiny}
\bibliographystyle{abbrv}
\bibliography{references}
\thispagestyle{empty}
\end{tiny}

\appendix
\section{Appendix}

\begin{lemma}[Chernoff bounds]
Let $X_1, \dots, X_n$ be independent Bernoulli random variables, and let $X=\sum_{i=1}^n X_i$. Then for any $0 < \delta < 1$ it holds that
\begin{enumerate}
\item For any $\mu \leqslant \Expec{X}$,
\begin{equation}\label{eq:cbl}
\Prob{X < (1-\delta) \mu} < e^{- \frac{\delta^2}{2} \mu}.
\end{equation}
\item For any $\mu \geqslant \Expec{X}$,
\begin{equation}\label{eq:cbr}
\Prob{X > (1+\delta) \mu} < e^{- \frac{\delta^2}{3} \mu}.
\end{equation}
\item For $\mu = \Expec{X}$,
\begin{equation}\label{eq:cbb}
\Prob{X \notin \left[ (1-\delta) \mu, \; (1+\delta) \mu \right] } < 2 e^{- \frac{\delta^2}{3} \mu}.
\end{equation}
\end{enumerate}
\end{lemma}

\end{document}